\documentstyle[prd,aps,twocolumn,epsf,axodraw]{revtex}
\def\br{\begin{eqnarray}}
\def\er{\end{eqnarray}}
\def\be{\begin{equation}}
\def\ee{\end{equation}}

\def\({\left(}
\def\){\right)}
\def\s{\sigma}

\begin{document}

\twocolumn[\hsize\textwidth\columnwidth\hsize\csname %%% TWO COLUMN
@twocolumnfalse\endcsname                            %%% TWO COLUMN
%
%\draft%
\title{
Hadronic physics in peripheral heavy ion collisions$^*$ }
\author{A.~A.~Natale  \\}
\address{
Instituto de F\'{\i}sica Te\'orica,
Universidade Estadual Paulista,
Rua Pamplona 145,
01405-900, S\~ao Paulo, SP,
Brazil}
%%%%
\date{\today}
\maketitle
%%%%
\begin{abstract}

We discuss the production of hadronic resonances in very peripheral heavy ion
collisions, where the ions collide with impact parameter larger than twice the
nuclear radius and remain intact after the collision. We compare the resonance
production through two-photon and double Pomeron exchange, showing that when we
impose the condition for a peripheral interaction the $\gamma \gamma$ process
dominates over the Pomeron interaction, due to the short range propagation of this last one.
We also discuss the observation of light resonances through the subprocess
$\gamma \gamma \rightarrow R \rightarrow \gamma \gamma $, which is a clean signal
for glueball candidates as well as one way to check the existence of a possible
scalar $\sigma$ meson.

\end{abstract}

\pacs{PACS: 25.75.-q, 27.75.Dw, 13.60.-r, 14.40.Cs}
\vskip 0.5cm]                           %two column

\section{Introduction}

Collisions at relativistic heavy ion colliders like the Relativistic
Heavy Ion Collider RHIC/Brookhaven and the Large Hadron Collider LHC/CERN
(operating in its heavy ion mode) are mainly devoted to the search of
the Quark Gluon Plasma. In addition to this important feature of
heavy-ion colliders, we will also have ultra-peripheral collisions with
impact parameter $b > 2 R_A$, where $R_A$ is the nuclear radius, and
where the ions remain intact after the collision.
\footnotetext{$^*$ Talk at the XXII Brazilian National Meeting on Particles
and Fields, S\~ao Louren\c co, Minas Gerais, 2001 }

These interactions will be mostly of electromagnetic origin: two-photon
($\gamma\gamma$) or photonuclear processes ($\gamma A$). Due to the very
strong photon field of each charge $Z$ accelerated ion, the photon
luminosity will be quite high. In the case of RHIC final states produced
in the two-photon process with an invariant mass up to a few GeV will
appear at large rates. Above this scale the photon luminosity drops
very fast. At LHC a final state with a mass almost two orders of
magnitude larger can still be produced at reasonable rates. The variety
of processes that can be studied in heavy ion peripheral collisions
have been extensively reviewed recently\cite{baur}.

The fact that hadronic resonances ($R$) could be produced at large rates
in peripheral heavy ion collisions was already discussed many years
ago\cite{bertulani}. Perhaps this may be one of the most interesting
studies to be performed at RHIC, because the machine will serve as a factory
of light hadrons in $\gamma\gamma$ and $\gamma A$ reactions. Vector resonances
will appear at huge rates in photonuclear reactions\cite{klein}, as well as scalar
and pseudoscalar resonances in two-photon processes\cite{natale}. We
will particularly focus our attention on the production of scalar and
pseudoscalar resonances through the $\gamma\gamma$ process.

Two-photon physics at $e^+ e^-$ colliders provided for a long time
a lot of information on hadronic resonances\cite{budnev}. The two-photon
process is very important because it involves the electromagnetic coupling of
the resonance, and its knowledge with high precision is very useful,
for instance, to unravel the possible amount of mixing in some
glueball candidates \cite{close}, complementing the information
obtained through the observation of hadronic decays. Another
interesting study is the possible production of a light scalar
meson ($\sigma$) whose existence has been discussed for several years\cite{torn}.
We stress again that the advantage of relativistic heavy ion colliders
is that the photon luminosity for two-photon physics is orders of magnitude
larger than the one at available $e^+ e^-$ machines.

\begin{figure}[htb]
\vskip 0.7in
%\centerline{
%\begin{picture}(300,100)(-15,-15)
\begin{picture}(290,10)(10,210)
 \Line(30,255)(120,255)
 \Line(30,250)(120,250)
 \Line(30,165)(120,165)
 \Line(30,160)(120,160)
 \GCirc(160,207.5){7}{0.5}
 \Photon(120,250)(155,212){2}{10}
 \Photon(120,165)(155,204){2}{10}
 \Photon(165,212)(210,240){2}{10}
 \Photon(165,204)(210,175){2}{10}
 \Line(120,255)(220,280)
 \Line(120,250)(220,275)
 \Line(120,165)(220,140)
 \Line(120,160)(220,135)
\end{picture}
\label{diagrama}
 \vskip 1.0in
  \caption{ Diagram for $\gamma \gamma $ fusion in a peripheral heavy-ion collision.
 The blob represents a continuum or resonant process leading to a two-photon final
 state.  }
\end{figure}
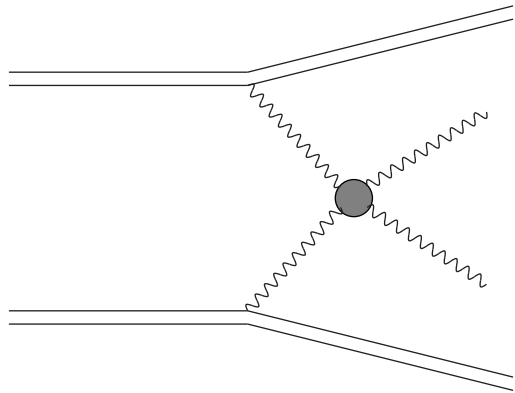

We will discuss the production of light hadronic resonances in
ultra-peripheral heavy ion collisions. We will show that
the process  $\gamma \gamma \rightarrow R \rightarrow \gamma \gamma $,
see Fig.(\ref{diagrama}),
can be observed for
many resonances above or at the same level of
the background. The main background is the continuum reaction
$\gamma \gamma \rightarrow \gamma\gamma$, this one will be discussed
as well as some other background contributions. Double Pomeron exchange
may also compete with the $\gamma \gamma$ physics, we will point out that
this contribution is not important for very heavy ions.

The distribution of
this review is the following: In Section II we present the distribution
functions for photons and Pomerons in the ion, and discuss some of the
approximations to obtain realistic cross sections. We compare $\gamma \gamma$
processes with the ones initiated by double Pomeron exchange. In Section
III we discuss the $\gamma \gamma \rightarrow R \rightarrow \gamma \gamma $
reaction and its background in the case of some glueball candidates and
in the case of a possible scalar $\sigma$ meson. Section IV contains our conclusions.

\section{Two-photon and double Pomeron exchange processes}

\subsection{Distribution functions}

The photon distribution in the nucleus can be described using the
equivalent-photon or Weizs\"{a}cker-Williams approximation in the
impact parameter space. Denoting by $F(x)dx$ the number of
photons carrying a fraction between $x$ and $x+dx$ of the total
momentum of a nucleus of charge $Ze$, we can define the
two-photon luminosity through
\begin{equation}
\frac{dL}{d\tau} = \int ^1 _\tau \frac{dx}{x} F(x) F(\tau/x),
\end{equation}

\noindent
 where $\tau = {\hat s}/s$, $\hat s$ is the square of
the center of mass (c.m.s.) system energy of the two photons and
$s$ of the ion-ion system. The total cross section of the process
$ AA \rightarrow AA \gamma \gamma $ is
\begin{equation}
\sigma (s) = \int d\tau \frac{dL}{d\tau} \hat \sigma(\hat s),
\label{sigfoton}
\end{equation}

\noindent
 where $ \hat \sigma(\hat s)$ is the cross-section of
the subprocess $\gamma \gamma \rightarrow X$.

There remains only to determine $F(x)$. In the literature there
are several approaches for doing so, and we choose the
conservative and more realistic photon distribution of
Ref.\cite{cahn}. Cahn and Jackson~\cite{cahn}, using a
prescription proposed by Baur~\cite{gbaur}, obtained a photon
distribution which is not factorizable. However, they were able
to give a fit for the differential luminosity which is quite
useful in practical calculations:
\begin{equation}
\frac{dL}{d\tau}=\left(\frac{Z^2 \alpha}{\pi}\right)^2
\frac{16}{3\tau} \xi (z),
 \label{dl1}
\end{equation}

\noindent
 where $z=2MR\sqrt{\tau}$, $M$ is the nucleus mass, $R$
its radius and $\xi(z)$ is given by
\begin{equation}
\xi(z)=\sum_{i=1}^{3} A_{i} e^{-b_{i}z},
 \label{dl2}
\end{equation}
which is a fit resulting from the numerical integration of the
photon distribution, accurate to $2\% $ or better for
$0.05<z<5.0$, and where $A_{1}=1.909$, $A_{2}=12.35$,
$A_{3}=46.28$, $b_{1}=2.566$, $b_{2}=4.948$, and $b_{3}=15.21$.
For $z<0.05$ we use the expression (see Ref.~\cite{cahn})
\begin{equation}
\frac{dL}{d\tau}=\left(\frac{Z^2 \alpha}{\pi}\right)^2
\frac{16}{3\tau}\left(\ln{(\frac{1.234}{z})}\right)^3 .
 \label{e5}
\end{equation}
The condition for realistic peripheral collisions ($b_{min} > R_1
+ R_2$) is present in the photon distributions showed above.

The processes that we shall discuss can also be
intermediated by the diffractive subprocess $P P \rightarrow
X$, where $P$ is the Pomeron.

In the case where the intermediary particles exchanged in the nucleus-nucleus
collisions are Pomerons instead of photons, we can follow closely the
work of M\"{u}ller and Schramm~\cite{muller} and make a
generalization of the equivalent photon approximation method to this new
situation. So the cross section for particle production via two Pomerons
exchange can be written as
\be
\sigma_{AA}^{PP} = \int dx_1 dx_2 f_P(x_1)f_P(x_2) \sigma_{PP}(s_{PP}),
\label{sechoque}
\ee

\noindent
where $f_P(x)$ is the distribution function that describe the probability for
finding a Pomeron in the nucleus with energy fraction $x$ and
$\sigma_{PP}(s_{PP})$ is the subprocess cross section
with energy squared $s_{PP}$. In the case of inclusive particle production
we use the form given by
Donnachie and Landshoff \cite{land3}
\be
f_P(x) = \frac{1}{4 \pi ^2 x} \int ^{-(xM)^2} _{-\infty} dt \,
 | \beta_{AP}(t)|^2 \, |D_P(t;s^\prime)|^2,
\label{in}
\ee

\noindent
where $D_P(t;s^\prime)$ is the Pomeron propagator\cite{land2}
\be
D_P(t;s) = \frac{(s/m^2)^{\alpha_P(t)-1}}{\sin ( \frac{1}{2} \pi \alpha _P(t))}
\exp{\left( - \frac{1}{2} i \pi \alpha _P(t) \right)}, \nonumber
\ee

\noindent
with $s$ the total squared c.m. energy.
The Regge trajectory obeyed by the Pomeron is
$\alpha _P(t) = 1 + \varepsilon + \alpha ^\prime _P t$, where $\varepsilon =
0.085$, $ \alpha ^\prime _P =0.25$ GeV$^{-2}$ and $t$ is a small exchanged
four-momentum square, $t= k^2 <<1$, so the Pomeron behaves like
a spin-one boson. The term in the denominator of the Pomeron propagator,
$ [\sin (\frac{1}{2} \pi \alpha _P(t))]^{-1}$, is the signature factor that
express the different properties of the Pomeron under C and P conjugation. At
very high c.m. energy this factor falls very rapidly with ${\mathbf{k}} ^2$,
whose exponential slope is given by $\alpha ^\prime _P \ln(s/m^2)$, $m$ is the
proton mass, and it is possible to neglect this ${\mathbf{k}}^2$ dependence,
\be
\sin \frac{1}{2} \pi (1+ \varepsilon - \alpha ^\prime _P {\mathbf{k}}^2)
\approx \cos (\frac{1}{2} \pi \varepsilon ) \approx 1. \nonumber
\ee

\noindent
If we define the Pomeron range parameter $r_0$ as
\be
r_0 ^2 = \alpha ^\prime _P \ln (s/m^2),
\label{r0}
\ee

\noindent
the Pomeron propagator can be written as
\be
|D_P(t=-{\mathbf{k}}^2;s)| = (s/m^2)^\varepsilon e^{-r_0^2 {\mathbf{k}}^2}.
\label{pomprop}
\ee

\noindent
Since we are interested in the spatial distribution of the virtual quanta in
the nuclear rest frame we are using $t=-{\mathbf{k}}^2$.

The nucleus-Pomeron coupling has the form \cite{land3}
\be
\beta_{AP}(t)= 3 A\beta_0 F_A(-t),
\nonumber
\ee

\noindent
where $\beta _0 = 1.8 $ GeV$^{-1}$ is the Pomeron-quark coupling, $A$ is the
atomic number of the colliding nucleus, and
$F_A(-t)$ is the nuclear form factor for which is usually
assumed a Gaussian expression (see, e.g., Drees et al. in \cite{papa})
\be
F_A(-t)= e^{t/2Q_0^2},
\label{fatf}
\ee

\noindent
where $Q_0=60$ MeV.

Performing the $t$ integration of the distribution function in Eq.(\ref{in}) we obtain
\begin{eqnarray}
f_P(x) &=& \frac{(3 A \beta_0)^2}{(2 \pi)^2 x}
\left( \frac{s^\prime}{m^2} \right)^{2 \varepsilon} \int ^{-(xM)^2}
_{-\infty} dt \,  e^{t/Q_0^2} \nonumber \\
       &=& \frac{(3 A \beta_0 Q_0)^2}{(2 \pi)^2x}
\left( \frac{s^\prime}{m^2} \right)^{2 \varepsilon} \exp \left[-\left(\frac{xM}{Q_0}\right)^2 \right].
\nonumber
\end{eqnarray}

The total cross section for a inclusive particle production is obtained with
the above distribution and also with the expression for the subprocess $PP
\rightarrow X$ as prescribed in Eq.(\ref{sechoque}).

\subsection{Is double Pomeron exchange a background for $\gamma \gamma$ processes?}

The double Pomeron exchange producing a final state $X$ have matrix elements
with the same angular structure as the $\gamma \gamma$ one.
This is easy to observe within the Donnachie and Landshoff model~\cite{land2}.
For instance, in this model to compute the cross section of the
subprocess $PP \rightarrow R$ it is
assumed that the Pomeron couples to the quarks of the resonance ($R$) like a
isoscalar photon \cite{land2}. This means that the subprocess cross section of $PP \rightarrow R$
can be obtained from suitable modifications on the cross section
for $\gamma \gamma \rightarrow R$, and it will differ only
by the coupling constant and a form factor describing the phenomenological Pomeron-quark
coupling. Therefore it is natural to ask if the Pomeron process is a
background to the two-photon process, and when it has to be added to the calculation
of a specific process.

In general the double Pomeron exchange is not a background for the
two-photon process and this is easy to understand. The Pomeron contrarily to the photon
does not propagate at large distances, actually its propagator has a
range parameter $r_0$ defined in Eq.(\ref{r0}). When we impose the condition
for ultra peripheral collisions ({\it i.e.} the nuclei do not physically collide)
the cross section diminishes considerably.

In Eq.(\ref{sechoque}) the cases where the two nuclei overlap are not
excluded. To enforce the realistic condition of a peripheral collision
it is necessary to perform the calculation taking into account the impact
parameter dependence, $b$. It is straightforward to verify that in the
collision of two identical nuclei the total cross section of Eq.(\ref{sechoque})
is modified to~\cite{muller}
\be
\frac{d^2 \sigma ^{PP \rightarrow X}_{AA}}{d^2 b} =
\frac{Q^{\prime 2}}{2 \pi} \, e^{-Q^{\prime 2} b^2/2} \, \sigma ^{PP}_{AA},
\label{dsin}
\ee

\noindent
where $(Q^{\prime })^{-2} = (Q_0)^{-2} + 2 r_0 ^2$. The total
cross section for inclusive processes is obtained after integration of
Eq.(\ref{dsin}) with the condition $b_{min} > 2R_A$ in the case of identical
ions.

Another way of to exclude events due to inelastic central collisions is
through the introduction of an absortion factor computed in the Glauber
approximation \cite{glauber}. This factor modifies the cross section
in the following form
\begin{eqnarray}
\frac{d \sigma^{gl}_{AA}}{d^2 b} &=&
\frac{d \sigma_{AA}^{{ PP} \rightarrow R}}{d^2 b} \, \exp \left[ -A^2 b
\sigma_0 \int \frac{dQ^2}{(2 \pi )^2} \, F_A^2(Q^2) \, e^{iQb}\right]
\nonumber \\
&=& \frac{d \sigma_{AA}^{{ PP} \rightarrow R}}{d^2 b} \, \exp
\left[ -A^2 b \sigma_0 \,\frac{Q_0^2}{4 \pi} \, e^{-Q_0^2 b^2/4} \right],
\label{gla}
\end{eqnarray}

\noindent
where $\sigma _0$ is the nucleon-nucleon total cross section, whose
value for the different energy domains that we shall consider is obtained
directly from the fit of Ref. \cite{caso}
\be
\sigma_0= X s^\epsilon + Y_1 s^{-\eta _1} +  Y_2 s^{-\eta _2},
\nonumber
\ee

\noindent
with $X = 18.256$, $Y_1 = 60.19$, $Y_2 = 33.43$, $\epsilon = 0.34$, $\eta _1 = 0.34$, $\eta _2 = 0.55$,
$F_A(Q^2) = e^{-Q^2/2Q^2_0}$ and we exemplified Eq.(\ref{gla}) for the case of
resonance production, {\it i.e.},  $\sigma_{AA}^{ PP \rightarrow R}$ is
the total cross section for the resonance production to be discussed in the
sequence. The integration in Eq.(\ref{gla}) is over all impact parameter
space

We compared the rates for double Pomeron exchange and two-photon
production of several final states like resonances, a pair of pions and a
hadron cluster of invariant mass $M_X$. The details of the calculations
can be found in Ref.\cite{eu} and here we will describe part of the
results of that work.

In Table I we compare the cross sections for resonance production
through the processes $ \gamma \gamma$,$P P \rightarrow R$.
\begin{table}[hb]
\center
\begin{tabular}{l c c c c c}
Meson & $M_R$ & $\Gamma_{(R \rightarrow \gamma \gamma)}$ &
{\it RHIC}$_{\gamma \gamma}$ & {\it LHC}$_{\gamma \gamma}$ &
  {\it RHIC}$_{\cal PP}$  \\
\hline
$\pi^0$ & 135  & $8 \times 10^{-3}$ &  7.1 & 40 &
0.05 \\
$\eta$ & 547  & 0.463 & 1.5 &17 &
 0.038 \\
$\eta ^\prime$ & 958  & 4.3 & 1.1 & 22  &
0.04 \\
$\eta_c$ &  2979 & 6.6 &$0.32 \times 10^{-2}$ & 0.5 &
$0.47 \times 10^{-4}$   \\
$\eta ^\prime _c$ & 3605  & 2.7  & $0.36 \times 10 ^{-3}$ & 0.1 &
 $0.34 \times 10^{-5}$  \\
$\eta _b$  & 9366  & 0.4 & $0.13 \times 10^{-7}$ & $0.37 \times 10^{-3}$  &
$0.11 \times 10^{-10}$ \\
\end{tabular}
\label{tab1}
\caption{Cross sections for resonance production through
photon-photon ($\gamma\gamma$) and double-Pomeron ($PP$) processes.
For $RHIC$, $\sqrt{s} = 200$ GeV/nucleon, we considered $^{238}$U ion and for
$LHC$, $ \sqrt{s} = 6 300$ GeV/nucleon, the nucleus is $^{206}$Pb. The
cross sections are in mbarn. Rates computed with the geometrical cut $b > 2 R_A$}
\end{table}
Table II shows the ratios of cross sections for diffractive resonance production
calculated with the Glauber absorption factor to the one with the geometrical
cut. The exclusion of central collisions through the Glauber absorption factor
is stronger than the one with the geometrical cut.
Table III shows the $\pi^0$ production for
different ions and at different energies. We observe that the rates for double Pomeron
exchange becomes closer to the two-photon rates when we go to lighter ions.
\begin{table}[hb]
\begin{center}
\begin{tabular}{l c c }
Meson & $\sigma_{AA}^{gl} / \sigma^{{\cal PP} \rightarrow R}_{AA}$ ({\it
LHC}) & $ \sigma_{AA}^{gl} / \sigma^{{\cal PP} \rightarrow R}_{AA}$ ({\it
RHIC})     \\ \hline
$\pi^0$ & $3.54 \times 10^{-3}$ & $1.5 \times 10^{-2}$ \\
$\eta $ & $3.58 \times 10^{-3}$ &  $1.47 \times 10^{-2}$ \\
$\eta ^\prime$ & $3.46 \times 10^{-3}$  & $1.5 \times 10^{-2}$ \\
$\eta _c$ &$3.47 \times 10^{-3}$  & $1.32 \times 10^{-2}$ \\
$\eta ^\prime _c$ & $3.61 \times 10^{-3}$ & $1.5 \times 10^{-2} $\\
$\eta _b$ &$3.5 \times 10^{-3}$ & $1.45 \times 10^{-2}$ \\
\end{tabular}
\caption{ Ratios of cross sections for diffractive resonance production
calculated with the Glauber absorption factor to the one with the geometrical
cut in
the collision of $^{238}$U for energies available
at RHIC ($\sqrt{s} = 200 $ GeV/nucleon), and collisions of $^{206}$Pb for
energies available at LHC ($\sqrt{s} = 6.300$ GeV/nucleon).}
\end{center}
\label{tab5}
\end{table}
\vskip -0.5cm
 The production of a cluster of particles through double Pomeron exchange is also
dominated by the $\gamma \gamma$ process (see Fig.(1) of Ref.\cite{eu}).

\begin{table}[hb]
\begin{center}
\begin{tabular}{l c c c c }
Nucleus & $\sqrt{s}$ & $\sigma ^{{\cal PP} \rightarrow R}_{AA}$  &
$\sigma^{gl^{\cal PP}}_{AA}$ & $\sigma _{\gamma \gamma }$   \\
\hline
Au ($A$=197) & 100  & 0.044 & $0.55 \times 10 ^{-3}$ &  2.4 \\
Ca ($A$=40) & 3 500 & 0.043 & $0.39 \times 10^{-3}$ & 0.14 \\
Si ($A$=28) & 200 & $0.34 \times 10^{-2}$ & $0.15 \times 10^{-3}$ & $0.69 \times 10^{-2}$ \\
Si ($A$=28) & 100 & $0.22 \times 10^{-2}$  & $0.12 \times 10^{-3}$ & $0.39 \times 10^{-2}$ \\
\end{tabular}
\caption{Cross section for $\pi^0$ production for
different ions and at different energies. The energies are in GeV/nucleon and
the cross sections in mbarn. $\sigma ^{PP \rightarrow R}$ is the cross section computed
with the geometrical cut and $\sigma ^{gl}$ is the one with the absorption
factor.}
\end{center}
\label{tab3}
\end{table}
In general we may say that for very heavy ions the double Pomeron exchange gives
cross sections one order of magnitude (or more) below the two-photon process. This
changes when we collide lighter ions. As we go down from large charge ions to smaller
ones the effect of Pomeron physics increases and it dominates
the electromagnetic physics, as happens in the proton case. The fact that
the Pomeron has a short range parameter is not the only fact that counts
when we analyze each specific process. It must be also remembered that
the Pomeron couples to light and heavy quarks differently.
Apart from kinematics the differences in the rates of resonance production
by photons and Pomerons increases between those resonances formed by
light and heavy quarks, as seen in Table I.

\section{The reaction $\gamma + \gamma \rightarrow \gamma + \gamma $}

\subsection{The continuum process}

We are particularly interested in the photon-photon scattering because it can be a
very clean signal for hadronic resonances like glueballs and the $\sigma$ meson.
On the other hand this scattering is important by itself, and
could probably be directly measured by the first time at RHIC,
as predicted in Ref.\cite{bertulani}.

The subprocess $\gamma \gamma \rightarrow \gamma \gamma $
up to energies of a few GeV is dominated by the continuous
fermion box diagram, and is a background for the resonant
$\gamma \gamma \rightarrow R \rightarrow \gamma \gamma$ process.
It was first calculated exactly by Karplus
and Neuman \cite{karplus} and De Tollis \cite{tollis2}.
There are sixteen helicity amplitudes for the process and,
due to symmetry properties, the number of independent
amplitudes will be only five, that may be
chosen to be $M_{++++}$, $M_{++--}$, $M_{+-+-}$, $M_{+--+}$ and
$M_{+++-}$. Where the $+$ or $-$ denotes the circular
polarization values $+1$ and $-1$. The remaining helicity
amplitudes may be obtained from parity and permutation symmetry.
Of these five helicity amplitudes, three are related by crossing,
hence it is sufficient to give just three, which are presented in
detail in Ref.\cite{we}.

The differential cross section of photon pair production from
photon fusion, i.e. the box diagram, is
\begin{equation}
\frac{d \sigma}{d \cos \theta }= \frac{1}{2 \pi} \frac{\alpha
^4}{s} (\sum_f q^2_f)^4 \sum |M|^2.
 \label{dcaixa}
\end{equation}

\noindent
  $\theta $ is the scattering angle,  $\alpha $ is the fine-structure
 constant, $q_f$ is the charge
 of the fermion in the loop and the sum is over the leptons $e$ and
 $\mu $ and the quarks $u$, $d$ and $s$, which are the relevant
 particles in the mass scale that we shall discuss. Another possible
 contribution to this continuum process comes from pion loops, which, apart
 from possible double counting, were shown to be negligible compared
 to the above one\cite{tollis3}.
The second sum is over the
 sixteen helicity  amplitudes, $M_{\lambda _1 \lambda_2 \lambda _3
 \lambda_4}$, where $\lambda _1$ and $\lambda _2$ correspond to
  polarizations of incoming photons and  $\lambda _3$ and $\lambda _4$
  for the outgoing photons.
The matrix elements summed over the final polarizations and
averaged over the initial polarizations is given by
\begin{eqnarray}
\sum |M|^2 &=& \frac{1}{2} \{ |M_{++++}|^2 + |M_{++--}|^2 +
|M_{+-+-}|^2  \nonumber \\
&+&  |M_{+--+}|^2 + 4 |M_{+++-}|^2 \}.
 \nonumber
\end{eqnarray}

We consider the scattering of light by light, that is, the
reaction $\gamma \gamma \rightarrow \gamma \gamma$, in Au-Au
collisions for energies available at RHIC, $\sqrt{s} = 200$
GeV/nucleon.  We checked our numerical code reproducing the many
results of the literature for the box subprocess, including
asymptotic expressions for the low and high energies compared to
the fermion mass present in the loop, and the peak value of the
cross section (see, for instance, Ref.\cite{landau}).

\begin{figure}[htb]
%\protect
\epsfxsize=.45\textwidth
\begin{center}
\leavevmode
 \epsfbox{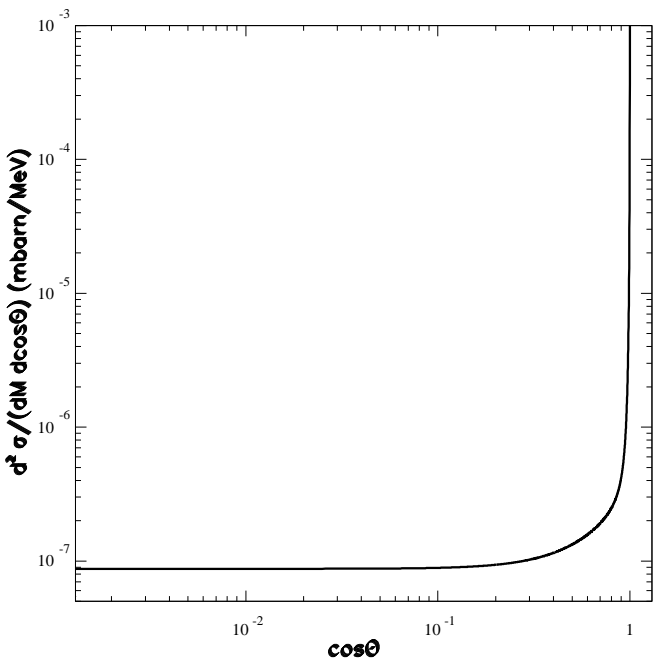}
\end{center}
\vskip -1.6cm
 \caption{Angular distribution of $ZZ \rightarrow ZZ\gamma\gamma$ scattering
 at an invariant mass of 500 MeV. The scattering  angle
 $\theta $ is in the photon pair center-of-mass system.}
  \label{dsigma_dcos}
\end{figure}

 In Fig.(\ref{dsigma_dcos}) the dependence of the ion-ion cross
section with the cosine of the scattering angle $\theta $, in the
two photon center-of-mass system, is shown for an invariant photon
pair mass equal to 500 MeV. It is possible to observe that the
cross section is strongly peaked in the backward direction, but is
relatively flat out to $\cos{\theta} \approx 0.4$, where it starts
rising very fast. It is symmetric with respect to $\theta $ and
the same behavior is present in the forward direction.

It is possible to observe that the cross sections
is strongly restricted when we introduce an angular cut.  We will
impose in all the calculations throughout this work a cut in the
scattering angle equal to $|\cos \theta | = 0.5$. This cut is
conservative, but it will make possible to compare the cross
section of the box diagram with rival processes, that will be
discussed in the following sections, as well as it is enough to
eliminate the effect of double bremsstrahlung (which dominates the
region of $| \cos{\theta} | \approx 1$). Finally, this kind of cut
is totally consistent with the requirements proposed in
Ref.\cite{klein}

 \vskip -1 cm
%\medskip
%\vskip .7cm

The fermions that contribute in the box diagram are the leptons
$e$ and $\mu$ and the quarks $u$, $d$ and $s$, which are
important for the mass range that we are interested (heavier
quarks will give insignificant contributions and the same is true
for the charged weak bosons). We assumed for their masses the
following values: $m_e = 0.5109$ MeV, $m_\mu = 105.6584$ MeV,
$m_u = 5$ MeV, $m_d = 9$ MeV and $m_s = 170$ MeV.
%\vskip -1.0cm

The electron gives the major contribution to the total result. The second
most important contribution is due to the muon, but it is at least one order of
 magnitude smaller than the electron one.
The $d$ and $s$ quark contribution (up to O(2 GeV)) are smaller
due to their masses and charges, because the process is
proportional to $(q_f ^2)^4$  where $q_f$ is their charge. Their
contribution is also insignificant compared to the electron
result.

As discussed in Ref.\cite{bertulani} the $\gamma \gamma$
scattering can indeed be measured in peripheral heavy ion
collisions. The cut in the angular distribution gives back to
back photons in the central rapidity region free of the
background. However, as we shall see in the sequence, there
are gaps where the $\gamma \gamma \rightarrow \gamma \gamma$
process is overwhelmed by the presence of resonances like the
$\eta$, $\eta\prime$ and others. Even the broad $\s$ resonance
could be of the order of the continuum process. Just to give one
idea of the number of events, with a luminosity of 2.0 $\times
10^{26}$ cm$^{-2}$s$^{-1}$ \cite{klein} and choosing a bin of
energy of 200 MeV, centered at the energy of 700 MeV (which is
free of any strong resonance decaying into two-photons), we have
1532 events/year assuming $100\%$ efficiency in the tagging of
the ions and photon detection.
%\vskip -2.0cm

\subsection{The process $\gamma \gamma \rightarrow R \rightarrow \gamma \gamma $: glueballs}

Photon pair production via the box diagram is a background to
$\gamma \gamma \rightarrow R \rightarrow \gamma \gamma $ process
(or vice versa), both have the same initial and final states, and
for this reason they can interfere one in another. Normally the
interference between a resonance and a continuum process is
unimportant, because on resonance the two are out of phase.

The total cross section for the elementary subprocess $\gamma
\gamma \rightarrow R \rightarrow \gamma \gamma $ assuming a
Breit-Wigner profile is
\begin{eqnarray}
\frac{d \sigma ^{\gamma \gamma}_{ZZ}}{dM} = 16 \pi \frac{dL}{dM}
\frac{\Gamma ^2 (R \rightarrow \gamma \gamma)} {(M^2-m_R^2)^2
+m_R^2 \Gamma^2_{total}},
 \label{dsigfoton}
\end{eqnarray}

 \noindent
 where $M$ is the energy of the photons created by
the collision of the ions. $\Gamma (R\rightarrow \gamma \gamma
)$($\equiv \Gamma_{\gamma \gamma}$) and $\Gamma_{total}$ are the
partial and total decay width of the
 resonance with mass $m_R$ in its rest frame.

\begin{figure}[htb]
\epsfxsize=.45\textwidth
\begin{center}
\leavevmode \epsfbox{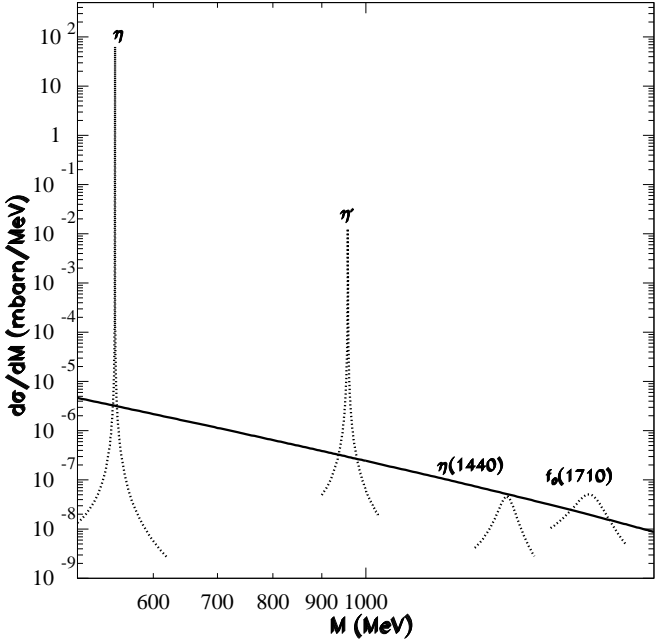}
\end{center}
\vskip -1.6cm
 \caption{Invariant mass distribution of photon production (with
the cut $|\cos{\theta}|<0.5$). The solid curve is for
 the box diagram, the dashed curves are due to the process
 $\gamma \gamma \rightarrow R \rightarrow \gamma \gamma $, where $R$ are
 the pseudoscalars resonances $\eta$ and $\eta^\prime$ and the glueballs
candidates $\eta(1410)$ and $f_0(1710)$.  }
  \label{glueball}
\end{figure}

 We are going to discuss only $J=0$ resonances made of quarks as
 well of gluons. The reaction $\gamma \gamma \rightarrow \pi^0
 \rightarrow \gamma \gamma$ was already discussed many years ago\cite{ora},
 where it was claimed that the interference vanishes. This result was
 criticized by De Tollis and Violini~\cite{tollis3}, affirming (correctly)
 that the interference exists. However, as we will discuss afterwards,
off resonance the interference is negligible. If the interference
is neglected, Eq.(\ref{dsigfoton}) can be used and we show in
Fig.(\ref{glueball}) the result for some resonance production
($\eta, \, \eta^\prime, \, \eta(1440), \, f_0(1710)$), whose
invariant mass of the produced photon pair is between 500 MeV and
2000 MeV. For comparison we also show the curve of the continuum
process. It is possible to see in that figure the well pronounced
peaks of the resonances $\eta$ and $\eta ^\prime$. We assumed for
their masses the values of 547.3 MeV and 957.78 MeV,
respectively, the $\eta$ total decay width is equal to 1.18 keV
and the $\eta^\prime$ one is equal to 0.203 MeV. Their partial
decay width into photons are 0.46 keV ($\eta$) and 4.06 keV
($\eta^\prime$).

\vskip -1.0cm

 We restrict the analysis to the $J=0$
glueballs candidates $\eta(1440)$ and $f_0(1710)$. For the
$\eta(1440)$ we used the mass and total decay width values of
Ref.\cite{pdg}, $m_R = 1405$ MeV, $\Gamma_{total} = 56$ MeV, for
the decay width into photons we use the value given in
Ref.\cite{close2}, $\Gamma_{\gamma \gamma} = 5.4$ keV. We see in
Fig.(\ref{glueball}) that the peak for this resonance is of the
same order of the continuum process. For the other glueball
candidate, $f_0(1710)$, the peak is clearly above the background.
For this one we assumed the values listed in Ref.\cite{pdg} of
mass and total width, $m_R = 1715$ MeV  $\Gamma_{total} = 125$
MeV, and for the two-photon decay width we adopted the value
encountered by the ALEPH Collab. \cite{aleph}, $\Gamma_{\gamma
\gamma} = 21.25$ keV. In all these cases the resonances can be
easily studied in peripheral heavy ion collisions.

Off resonance we can expect a negligible contribution for the
process $\gamma \gamma \rightarrow R \rightarrow \gamma \gamma $
and consequently the same for its interference with the continuum
process. However, it is instructive to present a more detailed
argument about why the interference can be neglected. In order to
do so we are obliged to introduce a model to calculate the
amplitudes for the process $\gamma \gamma \rightarrow R
\rightarrow \gamma \gamma $. These amplitudes will be computed
with the help of an effective Lagrangian for the pseudoscalar
interaction with photons (the scalar case will be discussed in
the next section), which is given by $g_p \varepsilon_{\lambda
\mu \nu \rho} F^{\lambda \mu} F^{\nu \rho} \Phi_p $, where $g_p$
is the coupling of the photons to the pseudoscalar field
$\Phi_p$, $\varepsilon_{\lambda \mu \nu \rho}$ is the
antisymmetric tensor and $F^{\lambda \mu}$ is the electromagnetic
field four-tensor.

The amplitudes for $\gamma \gamma \rightarrow \gamma \gamma $
intermediated by a pseudoscalar hadronic resonance are
~\cite{tollis3}:
\begin{eqnarray}
M_{++++} &=& \frac{2 \pi}{\alpha^2} F(\lambda r_t),
 \nonumber \\
M_{+-+-} &=& \frac{2 \pi}{\alpha^2} F(\lambda t_t),
 \nonumber \\
M_{+--+} &=& \frac{2 \pi}{\alpha^2} F(\lambda s_t),
 \nonumber \\
 M_{+++-} &=& 0,
  \nonumber \\
 M_{++--} &=& - \frac{2 \pi}{\alpha ^2} \{ F(\lambda s_t) + F(\lambda
 t_t) + F(\lambda r_t) \} ,
\label{pseudoescalar}
\end{eqnarray}

\noindent
 where $\alpha $ is the fine-structure constant, $\lambda
= (m_f/m_R)^2$, and
\begin{eqnarray}
F(x) = 16 x^2\frac{\Gamma_{\gamma \gamma}}{m_R} \left( 4 x - 1 + i
\frac{\Gamma _{total}}{m_R} \right)^{-1}.
 \label{fun_res}
\end{eqnarray}

The presence of the fine-structure constant in
 Eq.(\ref{pseudoescalar}) is a consequence of the fact that the
 amplitudes $M$ in these equations will be used in
 Eq.(\ref{dcaixa}), so it is necessary to get the correct
 dependence of the partial cross section with this constant.

A numerical evaluation of the cross section using
Eq.(\ref{pseudoescalar}) (for the same resonances present in
Fig.(\ref{glueball})) shows a totally negligible effect off
resonance in comparison with the box contribution. On resonance
the two processes are out of phase and the interference is
absent. We can now proceed with an argument showing that the
interference is not important. Let us assume that off resonance
the processes are in phase, and for a moment we forget the $t$
and $u$ channels contribution in Eq.(\ref{pseudoescalar}). The
$s$ channel contribution can be written as $M_2/(s - m_R^2 + i
\Gamma_R m_R)$, and denoting the continuum contribution by $M_1$
we can write the following interference term
\begin{eqnarray}
& & 2 \frac{s}{(s - m^2_R)^2 + \Gamma^2_R m^2_R} [ ( Re M_ 1 Re
M_2 + Im M_1 Im M_2 )
 \nonumber \\
  & & (s - m^2_R) + ( Re M_1 Im M_ 2 - Im M_1 Re M_ 2
) \Gamma_R m_R ].
 \nonumber
\end{eqnarray}

\noindent
 We can verify that the term proportional to $ s -
m^2_R$ integrates to zero when integrated in a bin centered at
$m^2_R$. With the second term the situation is different. If $Im
M_1$ or $Im M_2 \neq 0$ (assuming $ ReM_1 $ and $Re M_2 \neq 0$)
then there is a nontrivial interference. However, since we are
dealing with $J=0$ amplitudes, the only nonvanishing helicity
amplitudes are those in which the initial helicities and the
final helicities are equal. Inspection of the $M_{++++}$ and
$M_{++--}$ amplitudes of the box diagram (in the limit $m_f \ll
m_R$) shows that they are purely real, and the same happens with
$M_2$ (obtained from the $s$ channel contribution of
Eq.(\ref{pseudoescalar})), resulting in a vanishing interference!

Of course, this analysis is model dependent. In particular, at
the quark level the coupling $g_p$ has to be substituted by a
triangle diagram, which may have a real as well as an imaginary
part (see, for instance, Ref.\cite{novaes}). However, this
coupling is real for heavy quarks and its imaginary part is quite
suppressed if the resonance couples mostly to light quarks ($m_f
\ll m_R$, what happens for the resonances that we are considering
in the case of the u and d quarks, for the s quark the suppression
is not so strong in the $\eta$ case, but we still have an extra
suppression due to its electric charge). Finally, the
interference does appear when we consider the amplitudes with the
resonance exchanged in the $t$ and $u$ channels, but it is easy
to see that they are kinematically suppressed and also
proportional to the small value of the total decay width. If the
total width is large (as we shall discuss in the next subsection)
the interference cannot be neglected. Although the above argument
has to rely on models for the low energy hadronic physics, we
believe that the direct comparison between the resonant and the
continuum processes, as presented in Fig.(\ref{glueball}), is
fairly representative of the actual result.

In Table IV we show the number of events above background for the
$\eta$, $\eta^\prime$ and  $f_0(1710)$ which, as shown in
Fig.(\ref{glueball}) are clearly above the box contribution. In
the case of the $f_0(1710)$, as well as in the $\sigma$ meson
case to be discussed in the next subsection, the decay of the
resonance into a pair of neutral pions is present. A pair of
neutral pions can also be produced in a continuous two photon
fusion process. The rates for all the reactions discussed in this
work can be modified if both neutral pions, no matter if they
come from a resonance or a continuous process, are misidentified
with photons. This accidental background can be easily isolated
measuring its invariant mass distribution, and making a cut that
discriminates a single photon coming from the processes that we
are studying, from one that produced two neutral mesons
(mostly pions) subsequently decaying into two photons. For
example, in the case of the sigma meson (see next section) each
one of the neutral pions from its much large hadronic decay
should be misidentified. These pions would produce pairs of
photons with a large opening angle $\phi$, where $\cos{(\phi /2)}
= \sqrt{1 - 4m_{\pi}^{2}/m_{\sigma}^2}$. However, the
calorimeters already in use in many experiments are able to
distinguish between these two and single photon events with high
efficiency (see, for instance, Ref.\cite{alde}).
\begin{table}[hb]
\begin{center}
\begin{tabular}{l c }
 particle & events/year \\
\hline
$\eta$  & $7.44 \times 10^{5}$  \\
$\eta^\prime$ & $2.67 \times 10^{4}$ \\
$f_0(1710)$ & $42$  \\
\end{tabular}
\vskip 0.3cm \caption{Number of events/year above background for
the $\eta$, $\eta^\prime$ and  $f_0(1710)$ resonances. }
\end{center}
\label{tab6}
\end{table}
There is also another accidental background which may cause
problems to the measurement of the resonance decay into photons.
This is the contribution from $\gamma-$Pomeron$\rightarrow {\rm
V} \rightarrow \gamma \gamma X$.  The vector mesons, V, are
produced with a $p_T$ distribution similar to the resonance
production and at higher rates than some of the processes $\gamma
\gamma \rightarrow R$, {\it e.g.}\, $\gamma-$Pomeron$\rightarrow
\omega$ rate is 10 Hz at RHIC\cite{kln}, three orders of magnitude
higher than for a similar mass meson of spin 0 or 2. The $\omega$
branching ratio to three photons is 8.5\%. If a small $p_T$
photon from this decay is undetected, one is left with a low
$p_T$ two-photon final state that could be taken for a lighter
resonance. At higher masses, one also has $\phi \rightarrow \eta
\gamma$, $\pi^0 \gamma$, $K_L K_S \rightarrow \gamma X$ as well
as $\gamma \gamma \rightarrow f_2(1270) \rightarrow \pi^0 \pi^0$
and possibly copious production of $\rho(1450)$ and $\rho(1700)$
by $\gamma-$Pomeron interactions. Clearly a full simulation of all
these background processes should be kept in mind when measuring
two-photon final states.

\subsection{The process $\gamma \gamma \rightarrow R \rightarrow \gamma \gamma $: the $\sigma $ meson}

The possible existence of light scalar mesons (with masses less
than about 1 GeV) has been a controversial subject for roughly
forty years. There are two aspects: the extraction of the scalar
properties from experiment and their underlying quark
substructure. Because the $J=0$ channels may contain strong
competing contributions, such resonances may not necessarily
dominate their amplitudes and could be hard to ``observe". In
such an instance their verification would be linked to the model
used to describe them.

Part of the motivation to study the two-photon final states in
peripheral heavy-ion collisions was exactly to verify if we can
observe such scalar mesons in its $\gamma\gamma$ decay. Although
this decay mode is quite rare, it has the advantage of not being
contaminated by the strong interaction of the hadronic final
states. In particular, it may allow to investigate the possible
existence of the sigma meson. This meson is expected to have a
mass between 400-1200 MeV and decay width between 300-500 MeV,
decaying predominantly into two pions. Of course, another decay
channel is into two photons, with the background discussed in
the beginning of this Section.

Recently the E791 Collaboration at Fermilab found a strong
experimental evidence for a light and broad scalar resonance,
that is, the sigma, in the $D^+ \rightarrow \pi ^- \pi ^+ \pi ^+$
decay \cite{e791}. The resonant amplitudes present in this decay
were analyzed using the relativistic Breit-Wigner function
 given by
\begin{eqnarray}
BW = \frac{1}{m^2 - m^2_0 + i m_0 \Gamma(m)},
 \nonumber
\end{eqnarray}

\noindent
 with
\begin{eqnarray}
\Gamma(m) = \Gamma_0 \frac{m_0}{m} \left( \frac{p^*}{p^*_0}
\right) ^{2J + 1} \frac{^{J}F^2(p^*)}{^{J}F^2(p^*_0)},
 \nonumber
\end{eqnarray}

\noindent
 where $m$ is the invariant mass of the two photons forming a
 spin-J resonance. The functions $^JF$ are the Blatt-Weisskopf
 damping factors \cite{blatt}: $^0F = 1$ for spin 0 particles,
 $^1F = 1/\sqrt{1 + (rp^*)^2}$ for spin 1 and $^2F =  1/\sqrt{9 +
 3 (rp^*)^2 + (rp^*)^4}$ for spin 2. The parameter $r$ is the
 radius of the resonance ($\approx 3$ fm) \cite{argus} and $p^* =
 p^*(M)$ the momentum of decay particles at mass $M$, measured in
 the resonance rest frame, $p^*_0 = p^*_0(m_R)$. The Dalitz-plot of
 the decay can hardly be fitted without a $0^{++}$ ($\sigma$) resonance.
The values of mass and total decay width found by the
collaboration with this procedure are $478 ^{+24}_{-23} \pm 17$
MeV and $324^{+42}_{-40} \pm 21$ MeV, respectively.

We will discuss if this resonance can be found in peripheral
heavy-ion collisions through the subprocess $\gamma \gamma
\rightarrow \sigma \rightarrow \gamma \gamma $. It is important
to note that all the values related to the $\sigma$, like mass or
partial widths, that can be found in the literature are very
different and model dependent. In particular, we find the result
of the E791 experiment very compelling and among all the
possibilities we will restrict ourselves to their range of mass
and total decay width, while we vary the partial width into two
photons. For the $\sigma$ decay width into two photons we assume the
values obtained by Pennington and Boglione, $3.8 \pm 1.5$ keV and
$4.7 \pm 1.5$ keV \cite{pennington}, and the value of $10 \pm 6$
keV \cite{courau}.

It has been verified in the case of $\pi \, \pi$ scattering that
the use of a constant total width in the $\sigma$ resonance shape
is not a good approximation\cite{schechter}. In our case we will
discuss the $\gamma \gamma \rightarrow \gamma \gamma$ process
above the two pions threshold where the peculiarities of the broad
resonance, basically due to the $\sigma$ decay into pions, are not
so important. Of course, another reason to stay above $300$ MeV
is that we are also far from the pion contribution to $\gamma
\gamma$ scattering. In any case we also computed the cross
section with a energy dependent total width $\Gamma(m) \simeq
\Gamma_0 \left( {p^*}/{p^*_0} \right) ^{2J + 1}$, which, as shown
by Jackson many years ago\cite{jackson}, is more appropriate for
a quite broad resonance. The net effect is a slight distortion of
the cross section shape with a small increase of the total cross
section. Since this one is a negligible effect compared to the
one that we will present in the sequence we do not shall consider
it again.

Off resonance we can expect a negligible contribution for the
process $\gamma \gamma \rightarrow R \rightarrow \gamma \gamma $
and consequently the same for its interference with the continuum
process. This is true if the resonance has a small total decay
width\cite{we}, but this is not the $\sigma$ case. To take into account
the interference we must make use of a model to calculate the helicity
amplitudes of the $\sigma$ meson exchange. Using the effective
lagrangian $g_s F^{\mu \nu} F_{\mu \nu} \Phi_s $,
where $g_s$ is the coupling of the photons to the scalar field
$\Phi_s$ and $F^{\mu \nu}$ is the electromagnetic field tensor
the following amplitudes comes out\cite{we,tollis3}:
\begin{eqnarray}
M_{++++} &=& - \frac{2 \pi}{\alpha^2} F(\lambda r_t),
 \nonumber \\
M_{+-+-} &=& - \frac{2 \pi}{\alpha^2} F(\lambda t_t),
 \nonumber \\
M_{+--+} &=& - \frac{2 \pi}{\alpha^2} F(\lambda s_t),
 \nonumber \\
 M_{+++-} &=& 0,
  \nonumber \\
 M_{++--} &=& - \frac{2 \pi}{\alpha ^2} \{ F(\lambda s_t) + F(\lambda
 t_t) + F(\lambda r_t) \} .
\label{escalar}
\end{eqnarray}
where $\alpha $ is the fine-structure constant, $\lambda
= (m_f/m_R)^2$, and $r_t$, $s_t$ and $t_t$ are related with the standard
 Mandelstam variables s, t, and u by
$r_t = \frac{1}{4} \frac{s}{m_f^2}$, $s_t = \frac{1}{4}
\frac{t}{m_f^2}$ and $t_t = \frac{1}{4} \frac{u}{m_f^2}$.

\begin{figure}[htb]
%\protect
\epsfxsize=.35\textwidth
\begin{center}
\leavevmode \epsfbox{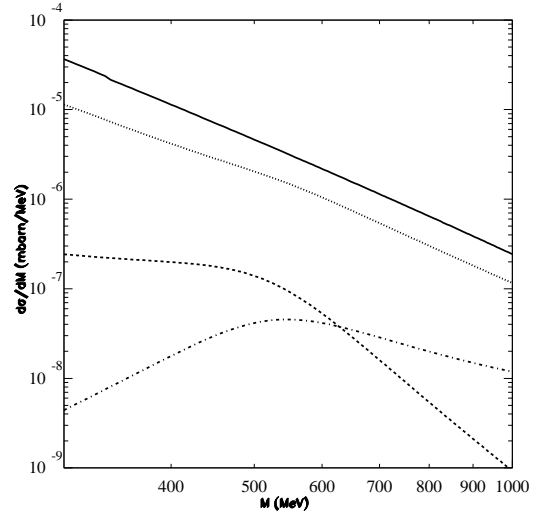}
\end{center}
\vskip -1.0cm
 \caption{ Invariant mass distribution of photon pair production.
 The solid curve is due to box diagram only, the dashed one is due to the process
 $\gamma \gamma \rightarrow \sigma \rightarrow \gamma \gamma$ in the
 Breit-Wigner approximation, the dash-dotted is the scalar contribution
 of Eq.(\ref{escalar}). In all cases $\Gamma_{\gamma\gamma} = 4.7$ keV,
 and the angular cut is equal to $-0.5 < \cos \theta < 0.5$ .  }
  \label{interferencia_escalar}
\end{figure}

These amplitudes and the ones describing the fermion (with mass
$m_f$) box diagram enter in the expression for the differential cross
section of photon pair production from photon fusion
${d \sigma}/{d \cos \theta }= ({1}/{2 \pi}) ({\alpha
^4}/{s}) (\sum_f q^2_f)^4 \sum |M|^2$,
to give the total cross section of Fig.(\ref{interferencia_escalar}).
We verified that the interference is destructive.
\vskip -1.0cm
\noindent

The effective Lagrangian model used to compute the $\sigma$
contribution to the photon pair production gives a larger
cross section than the calculation with the Breit-Wigner approximation
at energies above $M \approx 600$ MeV. It is dominated
by the $s$ channel contribution. We consider the Breit-Wigner result as the
best signal representation for the resonant process because we are
using the E791 data and this one was fitted by a Breit-Wigner profile.
The effective Lagrangian gives a nonunitary amplitude that
overestimates the sigma production above $600$ MeV and shows
the model dependence in the $\sigma$ analysis that we commented before.
The Breit-Wigner profile is not a bad
approximation as long as we stay above the two pions threshold and in the following
we assume that the signal is giving by it
(the dashed curve of Fig.(\ref{interferencia_escalar})) and the background is giving by
the box diagram result (the solid curve of Fig.(\ref{interferencia_escalar})). Note that, due to the
destructive interference, the actual measurement will give a curve
below the solid curve of Fig.(\ref{interferencia_escalar}).

From the experimental point of view we would say that the reaction
$\gamma \gamma \rightarrow \gamma \gamma$ has to be observed and
any deviation from the continuum process must be carefully
modeled until a final understanding comes out, with the
advantage that the final state is not strongly interacting.
Note that in this modeling the $\eta$ meson will contribute to
$\gamma \gamma \rightarrow \gamma \gamma$ in a small region of
momentum\cite{we}, even so it has to be subtracted in order to
extract the complete $\sigma$ signal.

\vskip -1.0cm
\begin{figure}[htb]
%\protect
\epsfxsize=.35\textwidth
\begin{center}
\leavevmode \epsfbox{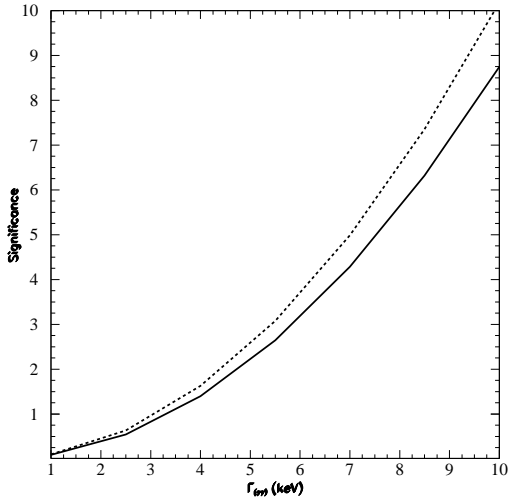}
\end{center}
\vskip -1.0cm
\caption{
Significance as a function of decay width into two photons,
 $\Gamma _{\gamma \gamma}$, for a sigma meson with mass equal to 478 MeV and
 total decay width of 324 MeV. The solid curve was obtained integrating
 the cross sections in the interval $438<M<519$ MeV, the dashed one in
 the interval $300<M<800$ MeV. The signal and background are giving by the
dashed (Breit-Wigner) and solid (box diagram) curves of
Fig.(\ref{interferencia_escalar}) respectively. }
  \label{significancia}
\end{figure}
\noindent

We changed the values of the $\sigma$ mass and total width around
the central ones reported by the E791 Collaboration. We do not
observed large variations in our result, but noticed that it is
quite sensitive to variations of the partial decay width into
photons. It is interesting to look at the values of the
significance which is written as $ {\cal
L}\sigma_{signal}/\sqrt{{\cal L}\sigma_{back}} $ and
characterizes the statistical deviation of the number of the
observed events from the predicted background. The significance
as a function of
the two photons decay width of the sigma meson,
with mass equal to 478 MeV and total decay width of 324 MeV, is
shown in Fig.(\ref{significancia}), were we used a luminosity of
${\cal L} = 2.0 \times 10^{ 26}$ cm$^{-2}$s$^{-1}$ at RHIC and
assumed one year of operation. The significance is above
$2\sigma$ $95\%$ confidence level limit for two photon decay
width greater than 4.7 keV, while for a $5\sigma$ discovery
criteria can be obtained with $\Gamma_{\gamma \gamma} > 7.5$ keV.
The numbers in Fig.(\ref{significancia}) were computed with the
signal given by the Breit-Wigner profile, and the background by
the pure box diagram. The solid curve was obtained integrating
the cross sections in the range of experimental mass uncertainty
$438<M<519$ MeV, while the dashed curve resulted
from the integration in the interval $300<M<800$ MeV. Note that
there is no reason, a priori, to restrict the
measurement to a small bin of energy. This choice will depend
heavily on the experimental conditions. Therefore, for
values of $\Gamma_{\gamma\gamma}$ already quoted in the
literature the sigma meson has a chance to be seen in its two
photon decay mode. The discovery limits discussed above refer
only to a statistical evaluation. Our work shows the importance
of the complete simulation of the signal and background including
an analysis of possible systematic errors that may decrease the
significance.

\section{Conclusions}

Peripheral collisions at relativistic heavy ion colliders provide
an arena for interesting studies of hadronic physics. One of the
possibilities is the observation of light hadronic resonances,
which will appear quite similarly to the two-photon hadronic
physics at $e^+e^-$ machines with the advantage of a huge photon
luminosity peaked at small energies~\cite{baur}. Due to this large
photon luminosity it will become possible to discover resonances
that couple very weakly to the photons~\cite{natale}.

The double Pomeron exchange may be a background for the
two-photon processes. We have shown that in general the
Pomeron contribution does not compete with the $\gamma \gamma $ one,
in the case of very heavy ions, after imposing the condition for
a ultra-peripheral collision. However it must be stressed that this
comparison has to be performed for each specific final state, due
to the different Pomeron coupling to several particles. If the
collision involves heavy ions and the final state we are looking
at does not have a large coupling to the Pomeron we can affirm
that double Pomeron exchange can be neglected in the evaluation
of production rates. In the case of light ions Pomeron processes
are competitive with the two-photons ones, and dominate the
cross sections for very light ions.

We discussed the peripheral reaction $A
A \rightarrow A A \gamma \gamma$. This process is important
because it may allow for the first time the observation of the
continuous subprocess $\gamma \gamma \rightarrow \gamma \gamma$
in a complete collider physics environment. This possibility only
arises due to enormous amount of photons carried by the ions at
the RHIC energies.

The continuous subprocess is described by a fermionic box diagram
calculated many years ago. We computed the peripheral heavy ion
production of a pair of photons, verifying which are the most
important contributions to the loop, which turned out to be the
electron at the energies that we are working, and established cuts
that not only ensure that the process is peripheral as well as
eliminate most of the background. After the cut is imposed we
still have thousands of photons pairs assuming 100$\%$ efficiency
in tagging the ions and detecting the photons.

The continuous $\gamma \gamma \rightarrow \gamma \gamma$ subprocess
has an interesting interplay with the one resulting from the
exchange of a resonance. We discuss the resonance production and
decay into a photon pair. This is a nice interaction to observe
because it involves only the electromagnetic couplings of the
resonance. Therefore, we may say that it is a clean signal of
resonances made of quarks (or gluons) and its measurement is
important because it complements the information obtained through
the observation of purely hadronic decays. It may also unravel
the possible amount of mixing in some glueball candidates
\cite{close}. We discuss the interference between these process
and compute the number of events for some specific cases.

The possibility of observing resonances that couple weakly to the
photons is exemplified with the $\sigma$ meson case. This meson,
whose existence has been for many years contradictory, gives a
small signal in the reaction $\gamma \gamma \rightarrow \sigma
\rightarrow \gamma \gamma$. However its effects may be seen
after one year of data acquisition, providing some clue about
this elusive resonance. Using values of mass, total and partial
widths currently assumed in the literature, we compute the full
cross section within a specific model and discuss the
significance of the events. Our work shows the importance of the
complete simulation of the signal and background of these
processes including an analysis of possible systematic errors,
indicating that two photon final states in peripheral collisions
can be observed and may provide a large amount of information
about the electromagnetic coupling of hadrons.

\section*{Acknowledgments}

Most of this work has been done in collaboration with C. G. Rold\~ao.
I would like to thank I.~Bediaga, C.~Dib, R.~Rosenfeld and
A.~Zimerman for many valuable discussions. This research was supported by
the Conselho Nacional de Desenvolvimento Cient\'{\i}fico e
Tecnol\'ogico (CNPq), by Fundac\~ao de Amparo \`a Pesquisa
do Estado de S\~ao Paulo (FAPESP) and by Programa
de Apoio a N\'ucleos de Excel\^encia (PRONEX).

\end{document}